\documentstyle[12pt,aasms4]{article}
\tightenlines
\begin{document}
\topmargin=-10mm
\baselineskip=18pt

\centerline{\bf Emission-Line Galaxies from the NICMOS/HST GRISM
Parallel Survey}
\bigskip
\author {Patrick J. McCarthy\altaffilmark{1}, Lin Yan\altaffilmark{1}, 
Wolfram Freudling\altaffilmark{2}, Harry I. Teplitz\altaffilmark{3,4},
           Eliot M. Malumuth\altaffilmark{3,6} }

\author {Ray J. Weymann\altaffilmark{1}, Matthew A. Malkan\altaffilmark{5}, 
Robert A. E. Fosbury\altaffilmark{2}, Jonathan P. Gardner\altaffilmark{3,4}}

\author { Lisa J. Storrie-Lombardi\altaffilmark{1},
Rodger I. Thompson\altaffilmark{7}, Robert E. Williams\altaffilmark{8}, Sara R. Heap\altaffilmark{3}}

\bigskip

\altaffiltext{1}{The Observatories of the Carnegie Institution of Washington, \\
                  813 Santa Barbara St., Pasadena, CA 91101}

\altaffiltext{2}{Space Telescope European Coordinating Facility,
                  Karl-Schwarzschild-Str. 2. D-85748 \\
                  Garching bei Munchen, Germany }

\altaffiltext{3}{NASA Goddard Space Flight Center, 
                  Code 681, Greenbelt, MD 20771 }

\altaffiltext{4}{NOAO Research Associate}

\altaffiltext{5}{Astronomy Department, University of California, 
                 Los Angeles, CA 90024-1562 }

\altaffiltext{6}{Raytheon STX Corp., Lanham, MD 20706}

\altaffiltext{7}{Steward Observatory, University of Arizona, 
                  Tucson, AZ 85720 }

\altaffiltext{8}{Space Telescope Science Institute, 
                  3700 San Martin Dr., Baltimore, MD 21218 }

\noindent {\it Subject headings:} cosmology: observations - galaxies: distances and redshifts,
evolution, starbursts - infrared: galaxies - techniques:
spectroscopic

\eject

\centerline{\bf ABSTRACT}

   We present the first results of a survey of random fields with the
slitless G141 ($\lambda_c = 1.5\mu$, $\Delta\lambda=0.8\mu$) grism on
NICMOS.  Approximately 64 square arcminutes have been observed at
intermediate and high galactic latitudes. The 3$\sigma$ limiting line
and continuum fluxes in each field vary from $7.5 \times 10^{-17}$ to
$1 \times 10^{-17}$ erg cm$^{-2}$ sec$^{-1}$ and from $H = 20$ to $22$,
respectively. Our median and area weighted $3\sigma $ limiting line
fluxes within a 4 pixel aperture are nearly identical at $4.1 \times
10^{-17}$ erg cm$^{-2}$ sec$^{-1}$ and are 60\% deeper than the deepest
narrow-band imaging surveys from the ground.  We have identified $33$
emission-line objects and derive their observed wavelengths, fluxes and
equivalent widths. We argue that the most likely line identification is
H$\alpha$ and that the redshift range probed is from $0.75$ to $1.9$.
The 2$\sigma$ rest-frame equivalent width limits range from 9\AA\ to
130\AA\, with an average of 40\AA.  The survey probes an effective
co-moving volume of $10^5~h_{50}^{-3}$~Mpc$^3$ for $q_0=0.5$. Our
derived co-moving number density of emission line galaxies in the range
$0.7 < z <  1.9$ is $3.3\times10^{-4}~h_{50}^{3}$~Mpc$^{-3}$, very
similar to that of the bright Lyman break objects at $z \sim 3$.  The
objects with detected emission-lines have a median F160W magnitude of
20.4 (Vega scale) and a median H$\alpha$ luminosity of $2.7 \times
10^{42}$ erg sec$^{-1}$. The implied star formation rates range from 1
to 324 M$_{\odot}$ yr$^{-1}$, with an average [NII]6583,6548 corrected
rate of 21 M$_{\odot}$ yr$^{-1}$ for H$_0=50$km/s/Mpc and $q_0=0.5$
(34~M$_{\odot}$ yr$^{-1}$ for $q_0=0.1$).

\eject

\bigskip
\section{Introduction}
\bigskip

Spectroscopic surveys of galaxies at $z < 1$  have reached a reasonable
level of maturity in recent years (e.g. CFRS, Lilly et al.  1996; Le
Fevre et al. 1996; CNOC, Yee et al. 1996). The use of specialized
selection techniques has led to a similar or greater gains at
redshifts of approximately 3 and larger (Steidel et al. 1996; Dickinson
1997; Hu, Cowie, \& McMahon 1998), with a few examples being identified
to $z \sim 5$ (Dey et al. 1998; Weymann et al. 1998; Spinrad et al. 1999).  However, to
build a coherent picture of galaxy formation and evolution we are
hindered by two problems. The first is that we know little of
properties of normal galaxies in the region between $ 1 < z < 2 $.
Secondly, most high redshift galaxies now known are selected at
rest-frame UV wavelengths, where the degree of extinction is
uncertain (e.g. Heckman et al. 1998).  The detection of heavily
dust-reddened galaxies at $z \sim 2 - 3$ in the sub-millimeter
wavelength (Smail, Ivison \&\ Blain 1997) suggests that high redshift
galaxies selected by UV colors represent only part of the population at
these early epochs.  At present, the estimates of the global star
formation rates are uncertain as they are based primarily on UV
selected samples (Pettini et al. 1997; Meurer et al. 1997).

   One approach to dealing with both the scarcity of spectral signatures in
the visible bandpass and the uncertainties arising from extinction is to observe in the
near-IR.  Until recently the available detectors and spectrometers have
made spectroscopic surveys in the near-IR impractical and so most
programs have been based on narrow band imaging ($\Delta \lambda \sim
1$\%) to search for strong emission-line objects, primarily in the K
window where H$\alpha$ can be detected for $ 1.9 < z < 2.6$.  Two
narrow band near infrared imaging surveys, covering several hundred
square arcminutes, were carried out with the Calar Alto 3.5m telescope
(Thompson, Mannucci \&\ Beckwith 1996; Beckwith et al. 1998; Mannucci
et al. 1998). The first Calar Alto survey targeted fields
containing known quasars or radio galaxies and the second the redshifts
of damped Ly$\alpha$ or metal line absorption systems. The first
survey discovered one emission line galaxy at $z = 2.43$ over an area
of 276 square arcminutes to a 3$\sigma$ flux limit of $3.4\times 10^{-16}$
erg cm$^{-2}$ s$^{-1}$, while the second survey found 18 emission line
galaxies (16 objects with detected H$\alpha$ and 2 with
[OII]3727) over an area of 163 square arcminutes to a
$3\sigma$ flux limit of $2\times 10^{-16}$ erg cm$^{-2}$s$^{-1}$. To date the
deepest narrow band near-IR imaging survey is that carried out by
Teplitz, Malkan \&\ McLean (1998) with NIRC on the Keck 10m telescope.
They found 13 emission line galaxies in an 11 square arcminute area to
a $3\sigma$ flux limit of $7\times 10^{-17}$ erg cm$^{-2}$s$^{-1}$.
The target fields of this survey were also selected by matching the
redshifts of quasar emission and metal absorption lines to their narrow
band filters. The space densities of strong H$\alpha$ emitting galaxies
inferred from these surveys span more than 2 orders of magnitude. The
large disparities in depths between the Calar Alto and Keck surveys and
the strong biases introduced by selecting redshifts matched to 
known absorption-line clouds make comparisons between the surveys and with other
measures of the average star formation rate non-trivial.

NICMOS offered a unique opportunity to address this problem with
multi-object spectroscopy.  The wide field camera on board NICMOS
(Thompson et al. 1998) is equipped with three grisms that allow slitless
spectroscopy over $0.75$ square arcminutes per exposure.  The extremely
low background at wavelengths below $1.9\mu$ allows for a sensitive low
resolution spectroscopic survey for H$\alpha$ and other emission-lines
through volumes that are an order of magnitude larger than those
available to narrow-band imaging surveys due to the large 
redshift range sampled by the grism.

Throughout this paper, H$_0=50$km/s/Mpc and $q_0=0.5$ are adopted 
unless explicitly indicated otherwise.

\bigskip \section{Observations} \bigskip

    All of the data presented here were obtained with Camera 3 on
NICMOS using the G141 grism and the F160W and F110W broad-band filters. 
The G141 spectral element covers the
wavelength range from $1.1\mu$ to $1.9\mu$ with a mean dispersion of $8.1
\times 10^{-3}\mu$ per pixel. The nominal resolving power of the grism
is 200. The actual resolution that is achieved is a function
of the apparent image size and  the camera 3 PSF. With the PSF present in the parallel
observations, the realized resolution is rarely better than R $\sim 150$. 
Small variations in the PSF
due to changes in the optical telescope assembly (e.g. breathing) and
longer term changes in the internal structure of NICMOS introduce small
variations in the maximum achievable resolution. As discussed in detail
below many of the objects detected in our survey are spatially resolved
and the resolution of their G141 spectra is determined by the image
size.

\subsection {Parallel observing mode and scheduling}

	All of the observations for this investigation were obtained
with HST operating in the parallel mode. While this allowed us to
collect far more data than would be possible in a single primary
program, it limited our ability to plan and execute the observations
in a manner that optimized the science return. Useful data for this
program first became available in October 1997 when the default
position of the pupil alignment mirror was moved from the camera 2
optimized position to the best reachable position for camera 3. After
that time camera 3 parallels were scheduled during much of the time for
which WFPC2 or STIS were the primary instruments. Most of the data in this
investigation was obtained as part of a public service parallel program
that was implemented and managed by the Space Telescope Science Institute.
The observing
algorithm was quite simple. In each full orbit one of the following
exposure sequences was selected: F160W and F110W imaging with exposure
times of either 512 or 382 seconds, F160W ($2 \times 256$ sec) and G141
($2 \times 896$ sec), or F110W ($2 \times 191$ seconds) and G096 ($2
\times 896$ sec). The selection of the exposure sequence for any given
orbit was nearly random, but was weighted in favor of the purely
imaging sequence for much of the October 1997 to March 1998 period. After
that time the balance was shifted towards spectroscopy with the grisms. All
of these observations were planned and implemented by the STScI staff,
with the guidance of Dr. John MacKenty.

    The array was read using the STEP64 read pattern. A $2^{''}$ dither
step was executed between each pair of direct or grism exposures.  This
dither step was made in the detector reference frame and was along
array rows. The dither step and orientation were fixed, resulting in
only two independent positions regardless of the number of exposures
for any given target. Small moves executed as part of the primary
observing plan introduced additional offsets for some of the
longer pointings.

  Approximately 10\%\  of our data were obtained as part of a crafted guest
observer (GO)
parallel program that was implemented for a portion of the February 1998 to
May 1998 period. The basic approach was similar to the public parallel
program, except that the dithering strategy was more complex.  For each
pointing we began with a number of F160W images with exposure times
selected to equal 15\%\ of the expected total duration of that
particular pointing. Dither moves were executed between each exposure,
but with step sizes and orientations that were non-redundant. The
remainder of the orbit and subsequent orbits were devoted to G141
exposures with a dither pattern matched to that of the direct images.
For a given exposure time the noise levels in these crafted observations were
$\sim 20$ to 30\% lower than those in the public parallels.

\bigskip
\subsection{Data processing and analysis}
\bigskip

\subsubsection {Two dimensional spectra}

We began our treatment of the data using the calibrated output from
the STScI data processing pipeline. The pipeline reduction consists of
dark current removal on a read-by-read basis, linearization, bad pixel
masking, and cosmic ray rejection.  At this stage the data remain in
the format of two dimensional, single frame images for each
exposure. The principal elements of the reduction process that are
accomplished by the pipeline to this point are the first order
bias and dark subtraction, linearization and CR rejection. Our
subsequent processing improves on the bias and dark 
subtraction, bad pixel masking and more complete rejection of particle
events. The calibrated grism exposures have not been corrected for the
flat field response and have not had the sky background removed. The
flat-field response is a function of wavelength for each pixel and so
depends on the location of each object, and will not be the same for
the object and the sky.  The flat-field correction is applied during
the 1-D spectral extraction.

We constructed a median sky frame for the G141 grism by combining the
calibrated exposures for approximately 200 frames, covering 80
independent positions on the sky.  We rejected observations made at
low galactic latitudes as well as those centered on bright objects or
crowded fields. This produced a high signal-to-noise uncorrected sky
frame for use in the sky subtraction for each exposure. The observed
background rates vary significantly from frame-to-frame depending on
the ecliptic latitude and longitude. We subtracted a scaled version of
the median sky from each calibrated 2-D spectral frame. The optimal
scaling for the sky was found by minimizing the RMS noise in two
object-free regions of each frame. Very low-spatial frequency
structure remained in the background of many frames, primarily due to
the shading effect in the NICMOS arrays. We removed a low order
surface from each frame.  This surface was derived from a second order
polynomial fit to an intermediate image constructed from the median of
each column.  This surface fitting was carried out separately in each
quadrant of the detector, allowing us to remove small
quadrant-to-quadrant bias variations.  The individual sky subtracted
and shading corrected frames were then shifted into registration,
using offsets derived from the accompanying direct images.  The
registered 2D spectra were then averaged with a bad pixel mask and
$3\sigma$ clipping applied to remove hot pixels and residual cosmic
rays.

The direct images that are used to calibrated the extracted one
dimensional spectra were processed and combined using the same
techniques as above, but without the sky subtraction and low order
surface fitting correction steps.

\subsubsection {1-D spectral extraction and calibration }

The extraction and calibration of the one-dimensional spectra was
carried out with the NICMOSlook software developed at ST-ECF
(Freudling et al. 1998).  The extraction was done in the following
war. First, the position and size of the objects previously
identified as spectral line objects were interactively marked on the
image display.  The extraction region of the spectrum was then
computed using transformations derived as part of the calibration
program.  The extracted spectra were corrected for the wavelength
dependent flat-field response on a pixel-by-pixel basis. For each
pixel, the quantum efficiency for the extracted wavelength was
estimated by interpolation from a series of narrow-band flat field
exposures. Fluxes were calibrated by applying throughput curves
previously derived from the observation of flux calibrators. The final
one dimensional spectra were obtained by adding the flux of all pixel
which correspond to each wavelength bin. For some objects close to one
of the edges of the grism images, only partial spectra could be
extracted.  Since there is a significant displacement between the
location of objects on the direct images and their dispersed spectrum
on the grism image, spectra for a few objects that fall outside the
area covered by the direct images were extracted. In that case, the
wavelength scale was estimated from the cut-off of the throughput
curves at high and low wavelengths which can easily be identified. For
such objects, the wavelength scale is uncertain by about 0.01$\mu$.

The rms accuracy of the wavelength calibration for G141 is about
$2\times 10^{-3}\mu$.  Repeated
calibration exposures have shown no changes larger than 3\% in the wavelength
zero point or 1.5\% in the dispersion. The
flux calibration is based on observations of both white dwarfs and
solar analogs during the NICMOS camera 3 campaigns.
The accuracy of the spectrophotometry is limited
by variations in the quantum efficiency within individual
pixels. We estimate that the flux scale for an individual spectrum is
accurate to about 10\%.

Spectra were also extracted in a parallel effort by members of the team
at the NASA Goddard Space Flight Center (GSFC)
 using software independent of the NICMOS-Look routines.  The
GSFC software followed a similar strategy to NICMOS-Look.  Spectral
positions on the chip were determined by the known offsets from the
object location in the direct image.  Extractions were performed on
individual (unregistered) frames first, with registration left to the
end.  Each spectrum was extracted in a weighted box, similar to the
procedure for the NICMOS-Look point source extraction.  The wavelength
solution was confirmed by independent analysis of the same calibration
data.  Flat-fielding was performed using the same data-cube of
narrow-band observations, interpolated to the wavelengths of the
extracted pixels, and the flux calibration was performed based on
independent analysis of the white dwarf observations.  Finally, spectra
were combined in wavelength space.  Emission-line wavelengths derived
from spectra extracted with the GSFC software were found to agree with
the NICMOSLook results to within one pixel (0.008$\mu$m).  Line fluxes
and equivalent widths were measured using Gaussian fits to the line and
polynomial fitting of the underlying continuum. The overall results of
the extraction and analysis using the GSFC and ECF/NICMOSLook software
were similar.

\bigskip
\section{Results}
\bigskip

\subsection {The properties of the data and the identification of emission-lines}

  There are several aspects of the data which negatively impact our
ability to identify genuine spectral features. Since we are observing
in the slitless mode, each object produces zero, first and second order
spectra. The first order spectra contain the useful data, the zero and
second orders are sources of confusion, the zero order
images as they can can be mistaken for emission features,
particularly when they fall on the first order continua of other objects. The zero order
images are slightly dispersed and for point sources often appear
bimodal. The displacement between the zero order images and the start
of the first order spectrum is $27^{''}$, or 135 pixels, a substantial
fraction of the array size.  For this reason the zero order images are
confined to roughly half of the detector area and in most cases they
can be identified by matching them with either the first order spectra
or with an object in the direct image.  There is a small portion of the
detector in which zero order images can appear without either first
order spectra or images in the direct frame. In addition to the
confusion caused by the zero order images and second order spectra,
there are artifacts associated with hysteresis in the HgCdTe
detectors.  Bright stars imaged with the broad-band filters just before
a grism observation will produce a low level after-image that is
displaced from the first order spectrum. Since the data are nearly
always taken in pairs of direct and grism exposures, these persistent
images appear in multiple exposures and can take on the appearance of
isolated emission-lines. Strong cosmic rays events have similar
properties, but the two position dithering sequence aids in their
identification and removal. Particularly strong particle events effect
more than one pixel and can produce amorphous clouds of electrons that
decay over time scales of an orbit. During the South Atlantic Anomaly
(SAA) passage the particle event rate is quite large and most NICMOS
images or spectra obtained at this time are not usable. The hysteresis
in the detectors often renders subsequent exposures useless as well.
For a significant number of our short pointings one of the two 896
second grism exposures has greatly enhanced noise from residual CR
events associated with the SAA passage.

   Given the wide variety of point-like features that appear in the two
dimensional grism images we have made our emission-line identifications
on the basis of visual inspection.  The eye can detect compact features
to quite low signal to noise levels. To convince ourselves that any
possible emission line was real we proceed to eliminate all possible
false signals, zero order images, persistent images, CRs etc. To
facilitate the detection of incompletely rejected cosmic rays and other
noise features we split each observation into two frames of equal
length and compared the position of possible emission features. Those
features that appeared in only one of the  data sets, or did not move
with the dither pattern, were rejected. In Figure 1 we show two pairs
of F160W/G141 frames and mark examples of each of the sources of
confusion as well as real emission features.

\subsection {Areal coverage and depth }

During the period from October 1997 to October 1998, 3500 camera 3
exposures were taken as part of the public and our crafted GO parallel
programs.  These contain a mix of direct exposures with the F160W and
F110W filters and spectroscopic exposures with the G141 and G096
grisms. A total of 456 pairs of F160W/G141 frames were taken during
this time.  These were distributed in 107 independent pointings with
exposure times ranging from 1790 seconds to 21,480 seconds. The G141
grism spectra can provide useful data for our purposes to quite low
galactic latitudes.  Rather than adopt a strict latitude cutoff we
chose to reduce all of the data and reject those with high stellar
densities. After some experience we rejected all fields with stellar
densities greater than 50 per square arcminute above a $10\sigma$
threshold in the F160W direct image. We are left with 85 pointings,
covering 64 square arcminutes.

  The final depth achieved varied considerably from field to field. For
the 1790 second exposures the noise level varied by a factor 3. Some of
this variation was due to genuine background variations, but much of it
was due to various non-Poisson sources of noise (e.g. SAA passage). For
longer pointings the depth increased roughly as the square root of the
exposure time. We define our limiting depths as $3\sigma$ within a 4
pixel (0.16 square arcseconds) aperture. This area is a reasonable
representation of the area of the camera 3 PSF at the focus achieved
outside of the campaign mode. In Figure 2 we show a histogram of the
$3\sigma$ 4-pixel limits for the G141 data.

\begin{figure}
\plotfiddle{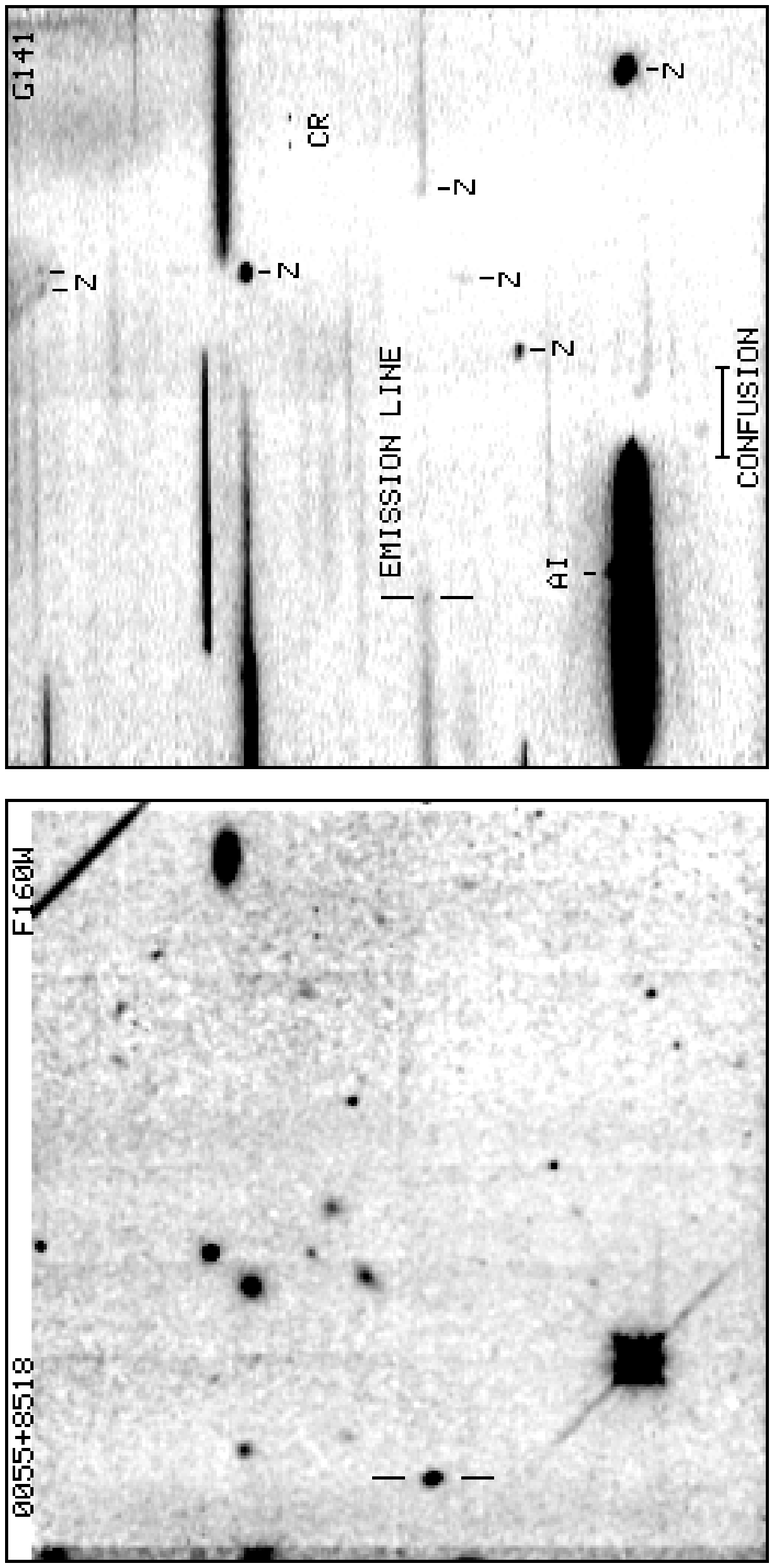}{7truecm}{-90}{50}{50}{-210}{380}
\plotfiddle{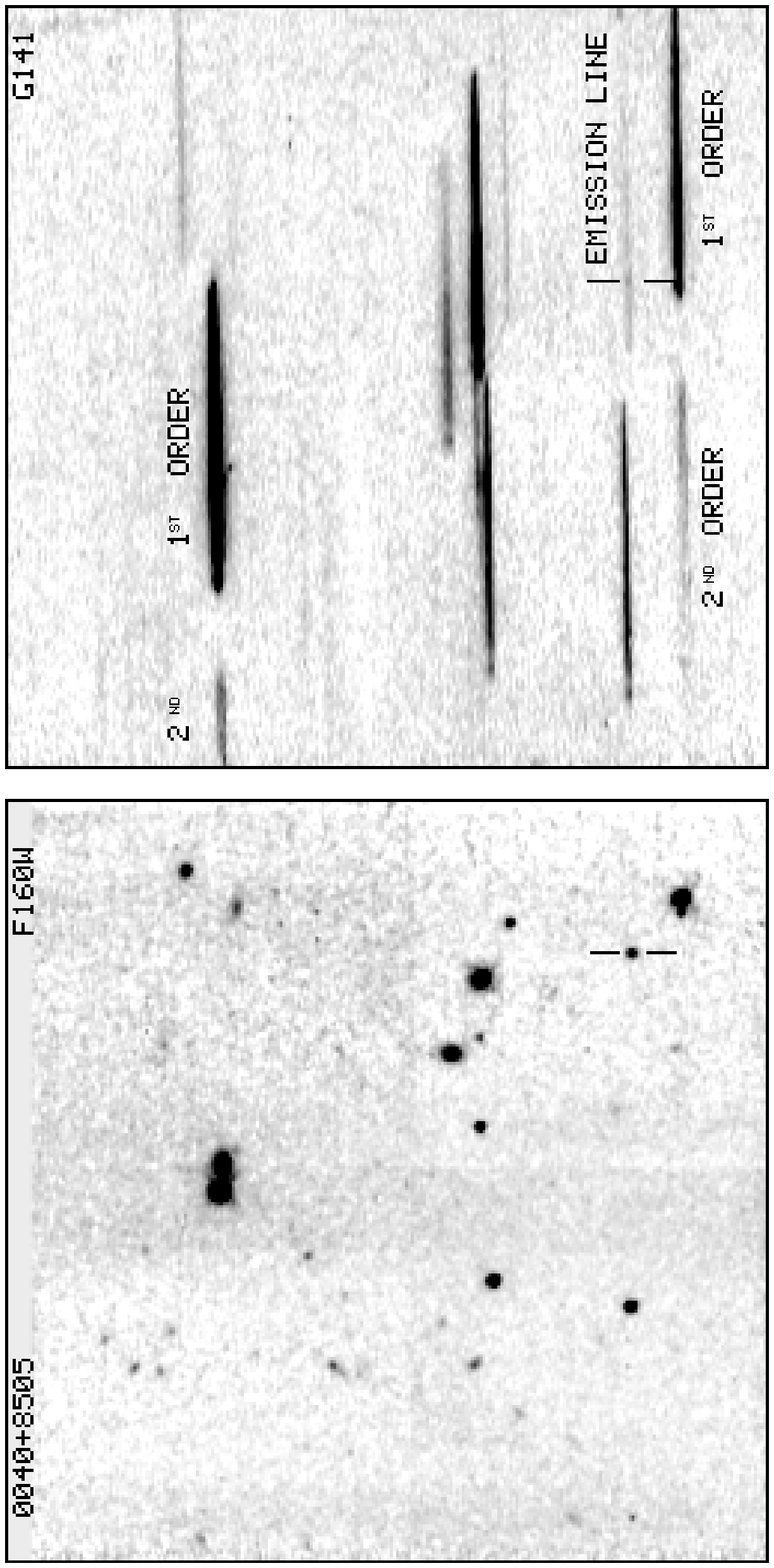}{0truecm}{-90}{50}{50}{-210}{240}
\caption{Two typical pairs of direct and 2-D grism images. Each field is
$51^{''} \times 51^{''}$ and oriented in arbitrary directions. In the
G141 frames we have marked zero order images (Z), residual
after images (AI), persistent cosmic rays
(CR), first and second order spectra and our identified emission-lines.
The region marked as ``confusion'' in the upper panel
identifies the portion of the detector
within which zero-order images neither have corresponding first-order
spectra within the G141 field of view or direct images in the F160W field.
We have marked the objects producing the emission lines in the F160W images.}
\end{figure}

\subsection {Detected Emission Features}

In Table 1 we list all of the fields for which we have reduced and
examined the G141 grism data. For each pointing we give the J2000 IAU
designation, the galactic latitude and the exposure times for the F160W
and G141 filters.

   In Figures 1a and 1b we show two typical pairs of direct and 2-D
grism images. These two grism images show many of the features
discussed in the previous section (zero order images, first and second
order spectra, after images) as well as one emission-line object each.
After careful examination of all of the spectra we assembled a catalog
of 33 emission-line detections.  Table 2 gives the field names, object
designation, J2000 coordinates, observed wavelengths, redshifts, line
fluxes, significance levels, observed equivalent widths and equivalent
width limits (2$\sigma$) for these 33 emission line galaxies.  We set a
rejection threshold at $3\sigma$, the derived significance levels of
our cataloged emission-lines are given in Table 2. These 33
emission-lines are distributed in 22 fields, 6 of which contain more
than one emission-line object. We do not convincingly detect more than
one line in any object.  All of lines are quite strong, in terms of
equivalent width.  This is expected given the low spectral resolution
of the grisms; weak lines against bright continua are not easily
detected.  The observed equivalent width limits reflect the
signal-to-noise ratios in the continua as well as the galaxy sizes.
The 2 $\sigma$ observed equivalent width limits are calculated as
follows: W$_{lim}= 2 \times ({s\over snr}) \times \sqrt{\rm N}$, with
$s$ the scale in \AA\ per pixel, $snr$ the signal-to-noise ratio per
pixel of the continuum at the emission line and $\rm N$ the number of
pixels under each line.  For lines that are not velocity resolved, the
apparent line width is the convolution of the instrumental spatial
resolution and the intrinsic size of the emission line region. For our
emission line galaxies, we take N to be 3 pixels.  The 2$\sigma$
rest-frame equivalent width limits range from 9\AA\ to 130\AA\ with an
average of 40\AA.  In Figure 3, we show direct images in the F160W band
($\sim$ H band) and 2D grism spectra for all 33 emission line galaxies.
Figure 4 presents the extracted one dimensional spectra of all 33
emission line galaxies.

\begin{figure}[h]
\plotfiddle{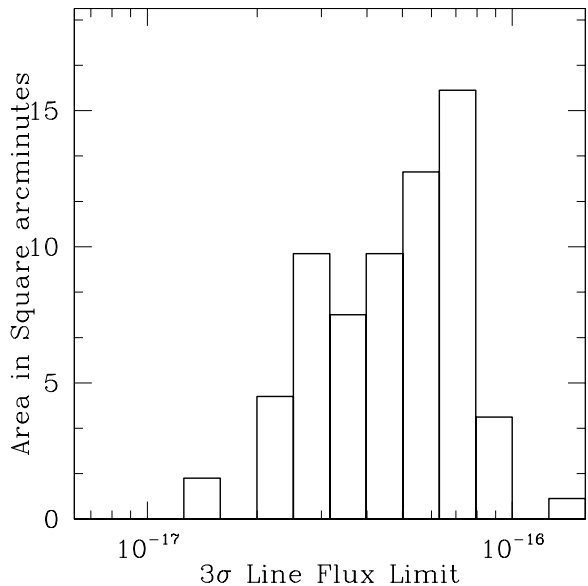}{7truecm}{0}{40}{40}{-150}{-70}
\caption{Histogram of the achieved $3\sigma$ 4-pixel noise levels in the
G141 data. Each pointing covers 0.75 square arcminutes. }
\end{figure}

\begin{figure}
\plotone{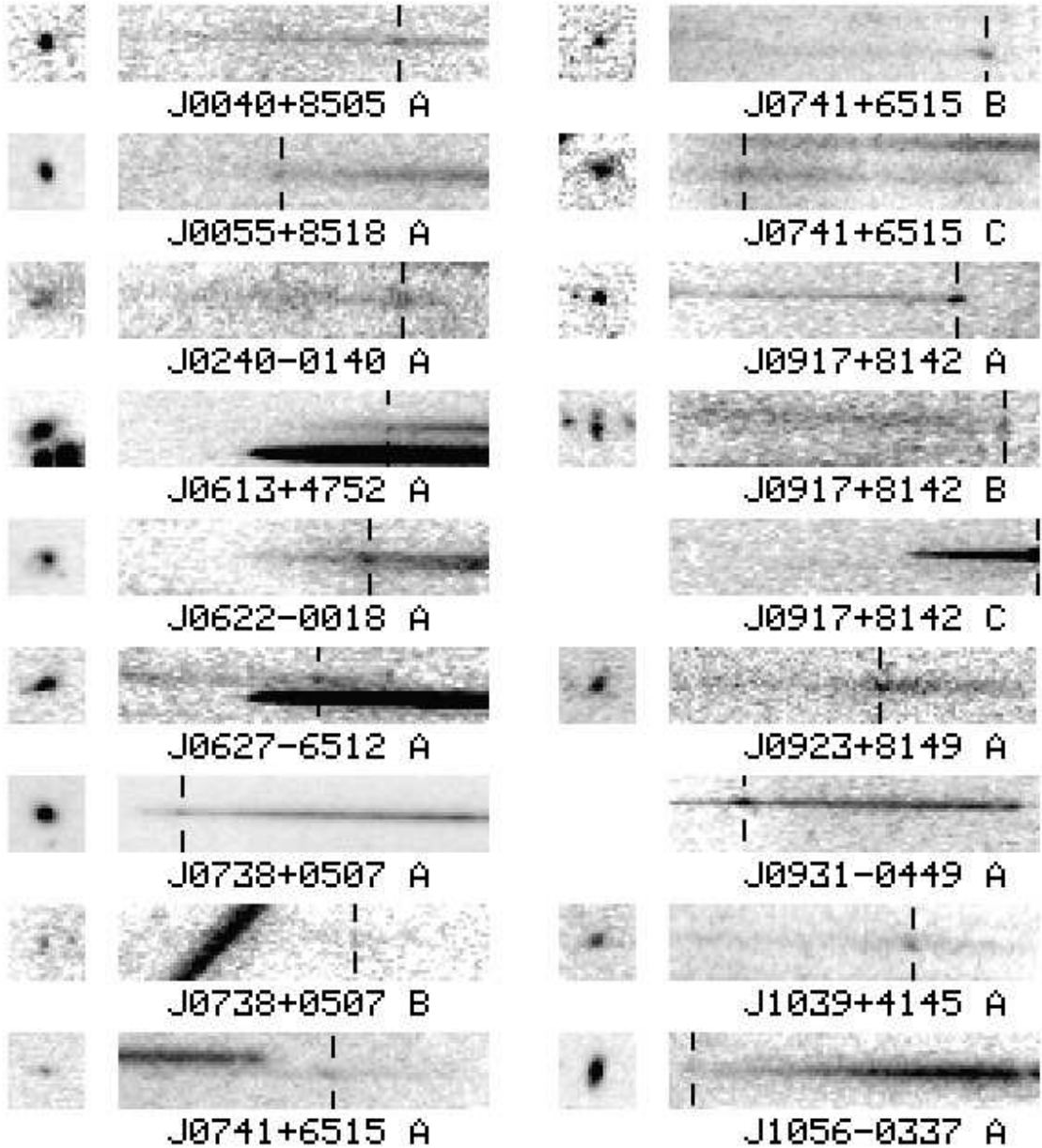}
\caption {A mosaic of F160W band direct images and
2D grism spectra of 33 emission line galaxies found in
our survey. The emission-lines in each object are marked with vertical
ticks. For a few of the objects the continuum images fall near the edge
or just off the edge of the F160W field of view. For these objects
(0917+8142C, 0931-0449A) we do not have F160W images.}
\end{figure}

\begin{figure}
\addtocounter{figure}{-1}
\plotone{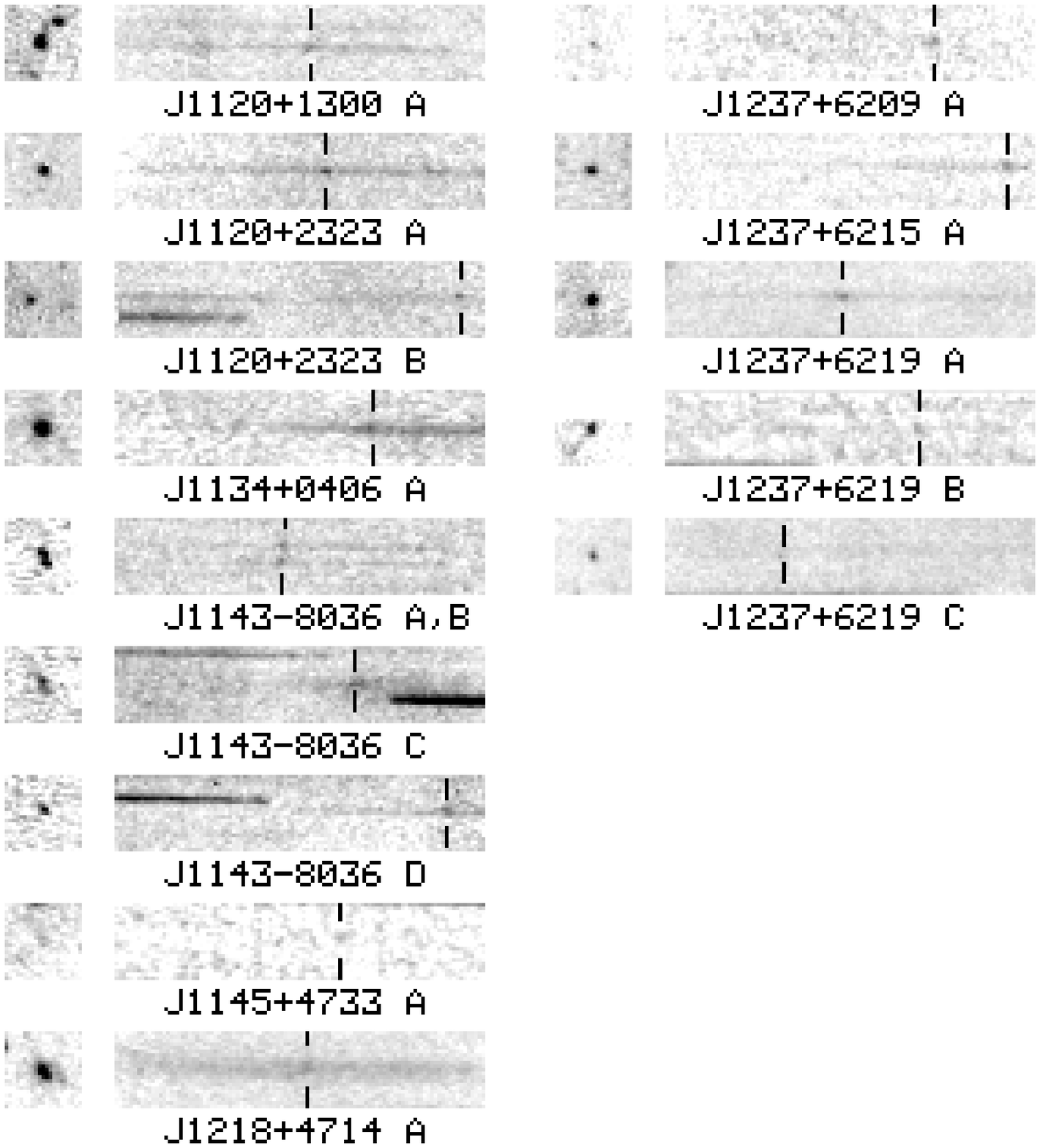}
\caption{continued.}
\end{figure}

\begin{figure}
\plotone{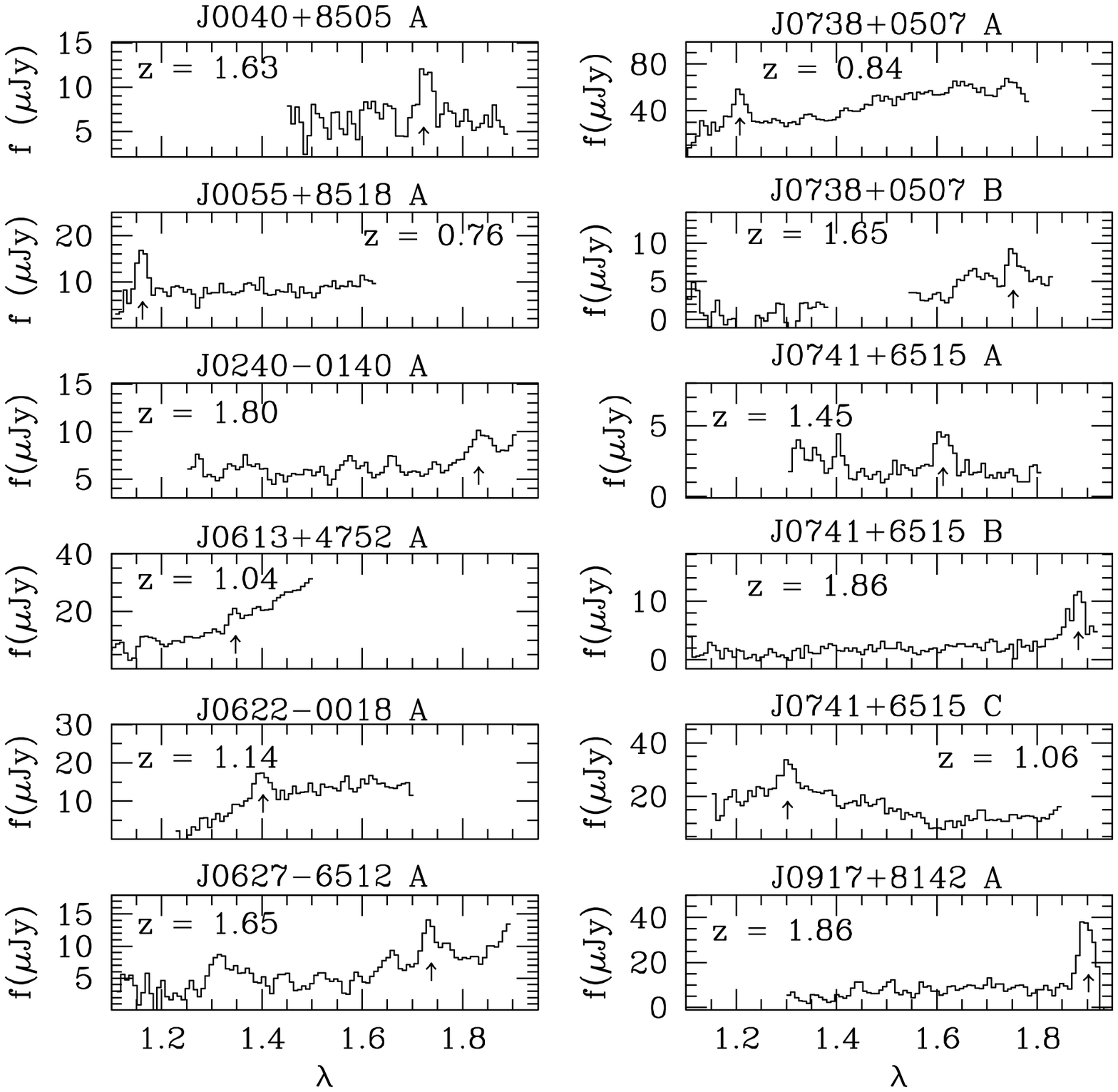}
\caption{ The extracted 1-D spectra of the emission line
galaxies. For each object we have marked our candidate emission-lines
with arrows below the line. We plot the entire range of the G141 grism
for each spectrum even though parts of the spectrum have fallen beyond
the field of view of the detector for several objects. The redshifts,
assuming an H$\alpha$ identification for the line are given for each
object. There are deficiencies in the background removal for several
objects that lead to significant uncertainties in the continuum level
(e.g. 1237+6215A)}
\end{figure}

\begin{figure}
\addtocounter{figure}{-1}
\plotone{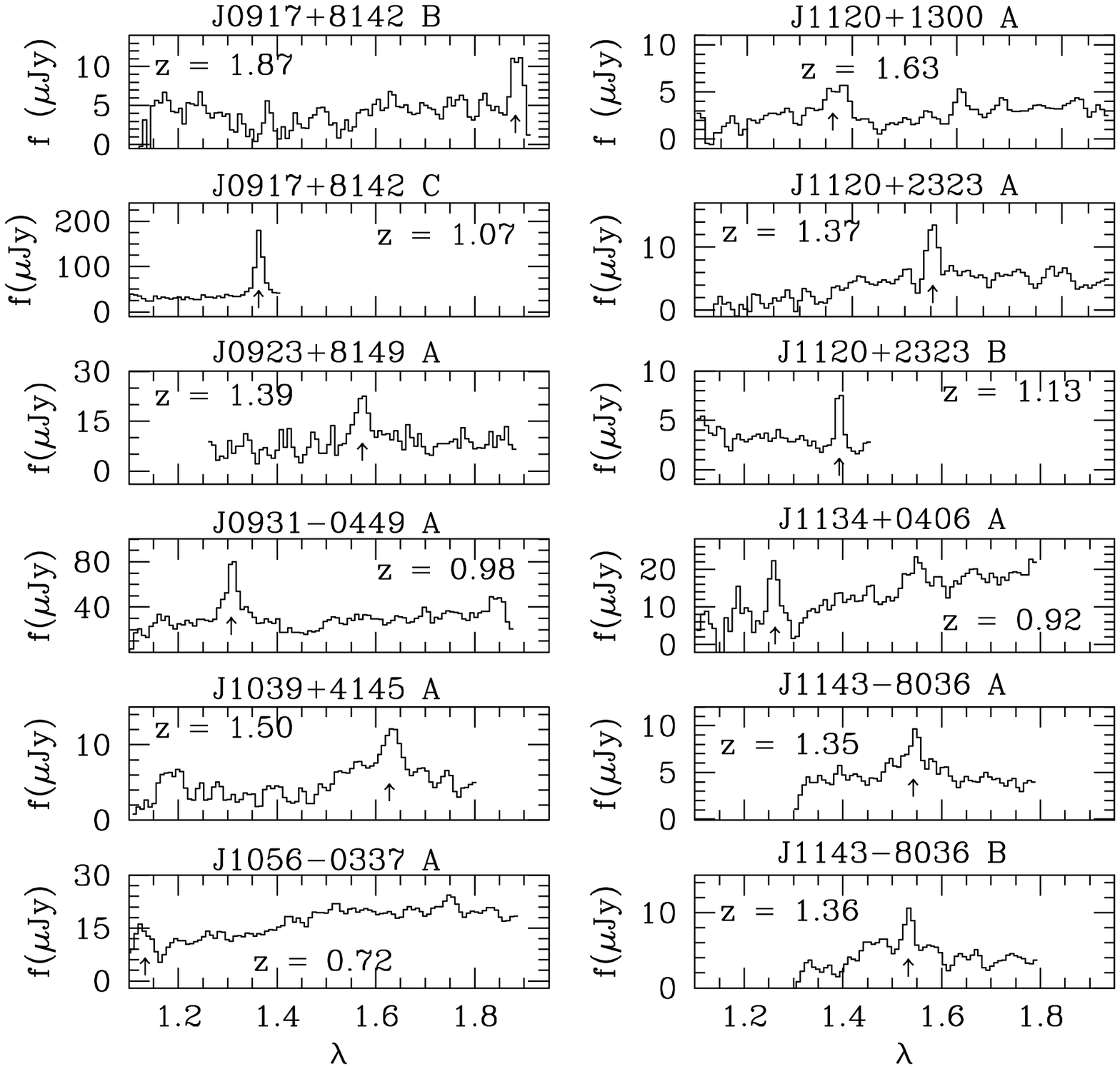}
\caption{continued.}
\end{figure}

\begin{figure}
\addtocounter{figure}{-1}
\plotone{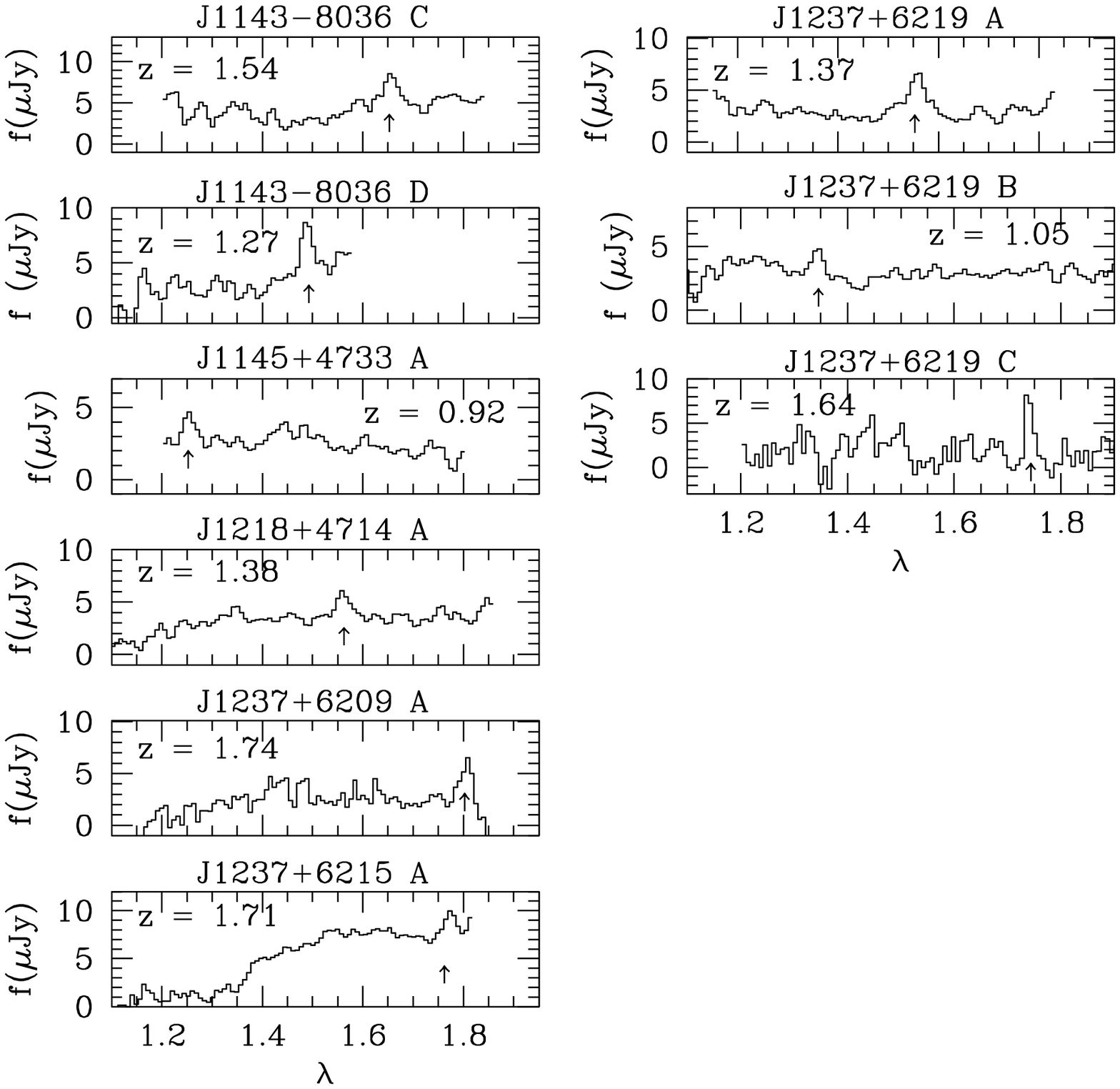}
\caption{continued.}
\end{figure}

We believe that most, if not all, of the emission lines that we have
detected are either H$\alpha$ or an unresolved blend of
H$\alpha$+[NII]6583/6548. In Table 3 we list the line luminosities, 
star formation rates derived from the Kennicutt (1984;
1992) relation, the apparent H-band magnitudes (on the Vega scale) and
J$-$H colors for the fields where the F110W observations were
available.  The identification of the emission-lines as H$\alpha$ is
based on a number of arguments.  First, the most plausible
identifications are H$\alpha$+[NII]6583/6548, [OIII]5007,4959,
H$\beta$4861 and [OII]3727. At the peak of the G141 transmission these
correspond to redshifts of 1.28, 2.0, 2.1 and 3.0, respectively.
Studies of nearby galaxies show that H$\alpha$+[NII] is usually the
strongest line, the second being [OII]$\lambda$3727 (Kennicutt 1992).
The equivalent width of H$\beta$ scales with the equivalent width of
H$\alpha$ as W(H$\beta$)$=0.15$W(H$\alpha$) - 4.  Even in the bluest
Magellanic irregulars and starburst galaxies, W(H$\beta$) is only $\sim
5$\AA. With a resolution of 5000 km sec$^{-1}$, it would be quite
difficult to detect such lines in our data.  Redshifts of $\sim 3$ are
highly unlikely given the median F160W magnitude of 20.5, as shown in
Figure 5.  The $z \sim 3$ Lyman break galaxies have typical F160W
magnitudes that are 2 magnitudes fainter (H $\sim 22.5$) (Giavalisco et
al.  1998).  The space density of our emission-line objects is
sufficiently large that they are unlikely to represent the extreme tail
of the luminosity distribution of the $z \sim 3$ Lyman break objects.
If not H$\alpha$, it is more likely that the lines that we see are
[OIII]5007,4959 rather than [OII]3727.  The [OIII]5007,4959 lines could
be marginally resolved in our spectra, but only if the objects are
spatially compact.  At $z \sim 1.5$, if we take an $R - H$ color of 3, the
median H magnitude of our objects corresponds to M$_R = -21.7$ for
H$_0=50$~km/s/Mpc, q$_0=0.5$, quite close to M$_R^{\ast}$ determined
from the Las Campanas Redshift Survey (Lin et al. 1996).  In Figure 6,
the left panel shows the distribution of H$\alpha$ line-luminosity; the
median line-luminosity is $2.5 \times 10^{42}$~erg s$^{-1}$, close to the value of
L$^{*}_{H\alpha}$ derived in local surveys (e.g. Gallego et al.  1995,
Tresse \& Maddox 1998). If the emission lines are actually [OIII]5007,4959,
the implied rest-frame H-band luminosities and emission-line
luminosities would be a factor of 2.5$-$3 larger and the likelihood of
such objects appearing in flux limited samples is suppressed by the
exponential cutoff in the luminosity function.  Strong [OIII]5007,4959
emission is more often identified with AGN than star-forming galaxies,
making it even less likely that [OIII] emitters at $z \sim 2$ dominate our
detections (see section 4).  Furthermore, in the redshift range between
1.6 and 1.8, both [OIII]5007,4959 and [OII]3727 are visible in the G141
passband.  Lines with longer rest-wavelengths are possible, but
unlikely identifications.  HeII10830 implies redshifts that are
implausibly low ($\sim 0.38$) and for most redshifts both of the
[SIII]9069,9545 lines would be within the passband.
Teplitz et al. (1999) have confirmed the identification of H$\alpha$ in two of
our objects (J0055+8518a, $z=0.76$, J0622-0018a, $z=1.12$) 
by detecting [OII]3727 at the appropriate redshifts in spectra taken 
with LRIS on the Keck 10m telescope.

H$\alpha$ is blended with [NII]6583,6548 in our spectra.  To correct
for the contribution from the [NII] doublet, we adopted
[NII]6583/H$\alpha = 0.3$, the average ratio for all types of galaxies
in the Gallego et al. (1997) objective-prism survey. Taking
[NII]6583/[NII]6548$ = 3$, we have $f({\rm H\alpha}) = 0.71\times
f(({\rm H\alpha + N[II]}))$. Thus the correction to the estimated star
formation rates in Table 3 is 29\%. We have not applied any extinction
corrections to either the H$\alpha$ luminosities or the star formation
rates.

\begin{figure}
\plotfiddle{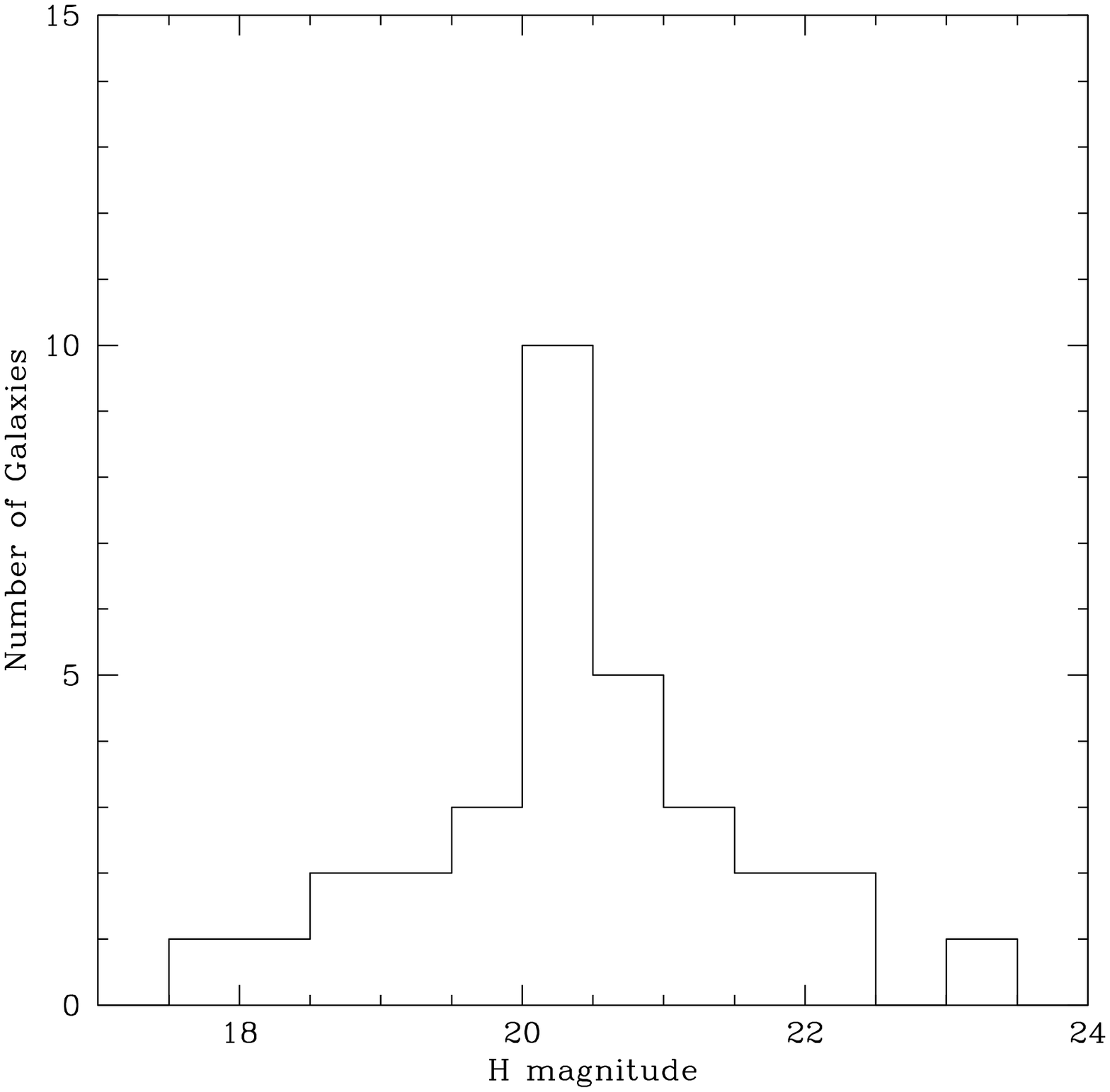}{7.5truecm}{0}{50}{50}{-150}{-95}
\caption{The distribution of apparent H magnitude of emission line galaxies.
The magnitudes are on the Johnson system.}
\end{figure}

\begin{figure}
\plotfiddle{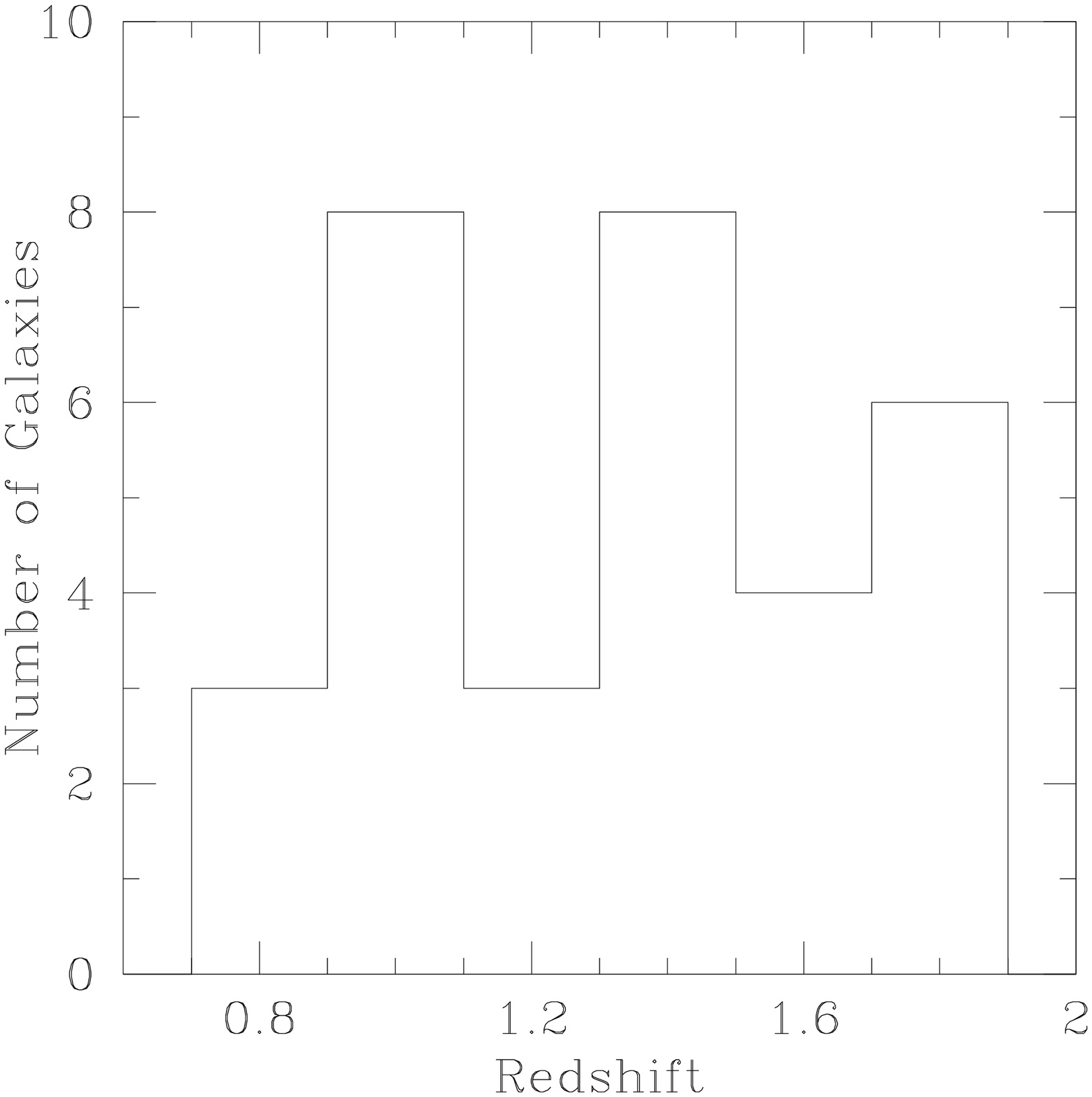}{6truecm}{0}{40}{40}{-250}{-110}
\plotfiddle{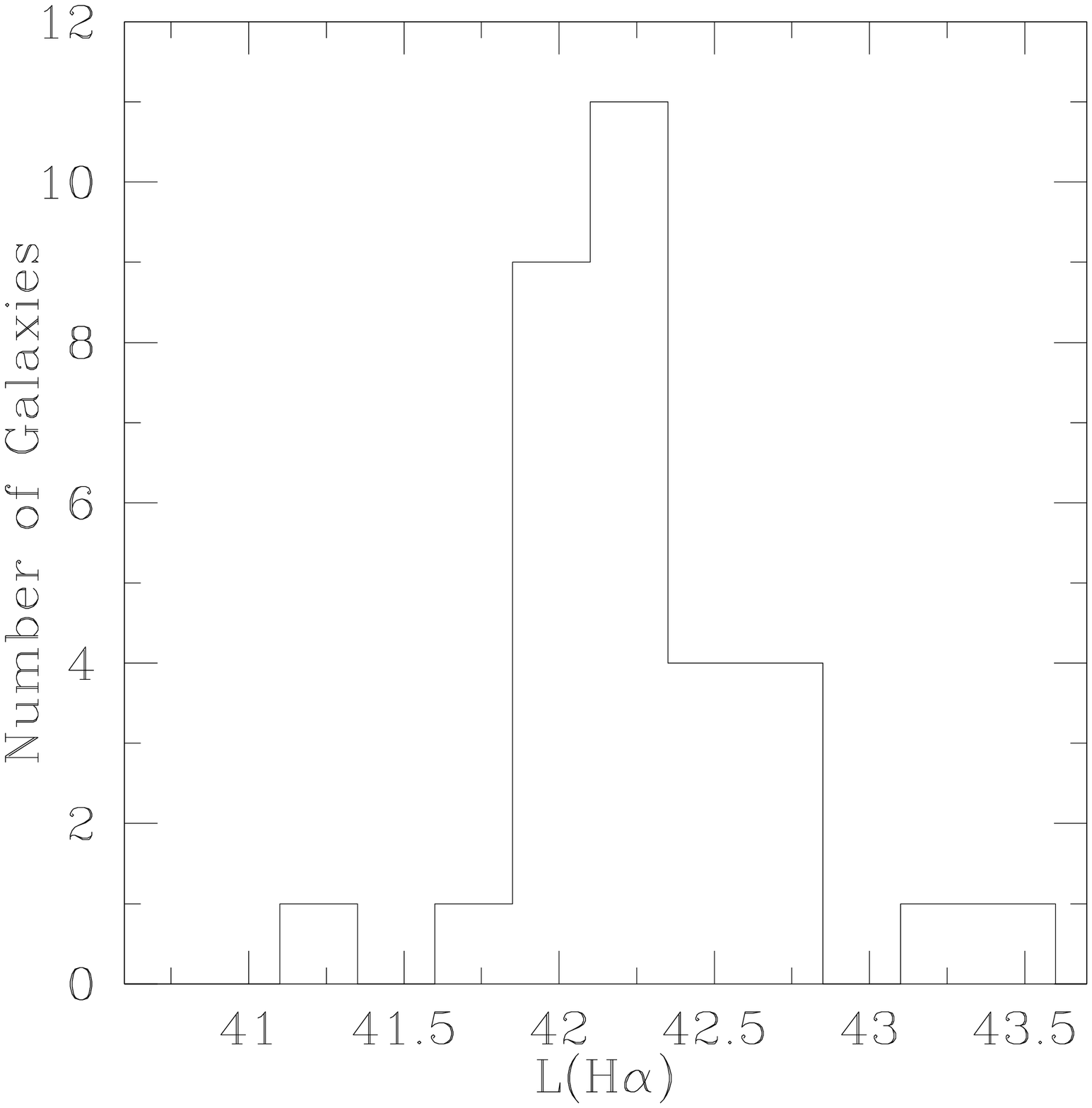}{0cm}{0.0}{40}{40}{-10}{-60}
\caption{The distributions of redshifts and H$\alpha$ line luminosities of the emission line galaxies.}
\end{figure}

\subsection {Notes on Individual Fields}

In Figure 3 we present cut-out images and 2-D spectra of all of our
emission-line candidates. Flux and wavelength calibrated 1-D spectra
are shown for the same objects in Figure 4. We designate each object by
its IAU J2000 field designation and a letter.  There are a few fields
that have noteworthy features and we detail these below.

\noindent 0040+8505 A: This is one of the weaker lines, in terms of W$_{\lambda}$, in
our sample. There are only two exposures of this field, but the emission feature
is seen in both frames. 

\noindent 0613+4752 A: The emission-line in this object is more compact than the
continuum. As in the case of 0040+8505 there are only two exposures for this
field, but the narrow emission feature is seen in both frames and moves with the
dither pattern.

\noindent 0622-0018 A: This low redshift object is detected in a very
low latitude field (b$ = -6^{\circ}$). The emission-line is spatially
extended and appears to follow the continuum distribution as revealed in
the F160W image.

\noindent 0738+0507 A: This is also a low latitude field and the object
in question is the brightest in our sample. The large equivalent width
and compact morphology of the emission-line suggests that this object
might be an AGN.

\noindent 0738+0507 B: The putative emission-line is quite faint and
diffuse. There are 6 exposures of this field and the emission feature
appears with equal strength in the two independent subsets of the
data.

\noindent 0741+6515 A: The emission-line in this object is spatially
resolved and its morphology is similar to that of the continuum shown
in the F160W image.

\noindent 0741+6515 B: The emission-line in this object is at the extreme
red end of the continuum spectrum and is spatially offset from the continuum.
The emission-feature is located in a portion of the field such that it cannot be
a zero-order artifact.

\noindent 0917+8142 A: This emission feature is also at the extreme red
end of the continuum. At the location of this feature there is the
possibility of zero order images. If this feature were a zero order
image, the object would be just off the edge of the F160W direct image.
The first order continuum of a zero order image at this location would
be partially on the G141 frame, but there is a confusing signal from a
bright galaxy at this location. 
The spatial coincidence between the continuum and
emission features led us to classify this as a real emission-line, although
this object should be considered less secure than the others.

\noindent 0917+8142 C: The putative emission-line in this object is at
the edge of the G141 field of view. The feature is only present in one
of the two dithered frames, it was moved off the detector by the
$2^{''}$ dither motion.  The object is also located just outside the
F160W field in both dither positions.  The great strength of this line
makes it a candidate AGN.

\noindent 1120+2323 A,B: The two emission-lines in this field are
extremely compact. Both emission features are seen in the two dither
positions.  For reasons that are not entirely clear, the focus in this
particular set of exposures was unusually good. The difference may have
arisen from breathing of the telescope optical assembly, or from
motions within the NICMOS dewar.

\noindent 1143-8036 A,B: This pair of emission-line galaxies appear to
have nearly identical redshifts in the 2-D G141 frame. The slightly
different locations in the F160W image result in different wavelength
scales, and hence different redshifts. The close proximity of the two
images make accurate centroiding of the F160W images difficult and this
introduces uncertainties into the wavelength calibrations.

\bigskip
\section{Discussion}
\bigskip

  This survey provides a unique window on galaxies at intermediate
redshifts. In so doing it provides us with a view of the bright end
of the H$\alpha$ luminosity function averaged over a cosmologically
significant volume. From these data we will derive an
H$\alpha$ luminosity function and a measure of the volume averaged
star formation rate that is fairly insensitive to the effects of
extinction. These analyses require careful treatment of the selection
efficiencies and biases inherent in a slitless low-resolution survey
and are presented in Yan et al. (1999). In the present
work we restrict ourselves to a first-order census of bright
emission-line objects and a brief comparison with other surveys in this
redshift range.

   There are a number of near-IR emission-line imaging surveys that are
either in progress or have been recently completed. In Table 4 we
compare the characteristics of our survey with several ground-based
programs. Clean comparisons are difficult since our survey is of a
different character and spans a very wide range of exposure times from
field to field.  In our deepest pointings we achieve flux limits that
are substantially deeper than the narrow-band imaging survey being
carried out by Teplitz et al. (1998) with the Keck Telescope. Our
median and area weighted sensitivities  are both $4.1 \times 10^{-17}$
erg cm$^{-2}$ sec$^{-1}$ (3$\sigma$, 4 pixels), $\sim 60$\% deeper than
the area-weighted detection limit reported by Teplitz et al. (1998).
The two principal differences between our survey and those listed in
Table 4 are the volumes probed and the choice of sight-lines. Because
our survey is spectroscopic we are able to probe large columns in the
$z$ direction. The comparable areal coverage of our survey then leads
to total co-moving volumes that are one to two orders of magnitude
larger than those surveyed by the narrow-band imaging programs.

Inspection of Table 4 reveals that most of the narrow-band imaging
surveys are targeted at lines of sight and redshifts that are {\it a priori} known to
contain either luminous galaxies or absorption line systems. The
motivation for this can be clearly seen in the results of the first
Calar Alto survey (Thompson et al. 1996). Surveys along sight-lines
without absorbers, or in completely random areas, yield few strong line
emitters in the volumes that can be probed by imaging with 1\% filters.
Thompson et al. detected a single object in 30,000~$h^{-3}_{50}$Mpc$^3$, 
implying a space density of $3 \times 10^{-5}~h^3_{50}$Mpc$^{-3}$.
The choice of AGN and/or absorber redshifts is effective in producing
detections, but it clearly impacts the inferred space density of line
emitters. Teplitz et al. (1998), targeting a mix of QSO and
absorber redshifts, report the highest space density, at
approximately $10^{-2}$~$h^3_{50}$Mpc$^{-3}$. This is quite different
from the results of the Thompson et al. (1996) survey, even when
allowing for the very different depths of the two surveys. 
The second Calar Alto survey (Mannucci et al. 1998) demonstrates the
effect of targeting metal line absorption systems, while Bechtold et
al. (1997) have achieved similar results using damped Ly$\alpha$
redshifts. 


Our H$\alpha$ survey probes an effective co-moving volume of $1.0\times
10^{5}~h_{50}^{-3}$~Mpc$^3$ ($2.1 \times
10^5~h_{50}^{-3}$~Mpc$^3$ for q$_0=0.1$).  Thus the co-moving number
density of emission-line galaxies averaged over our volume is
$3.3\times10^{-4}~h^3_{50}$~Mpc$^3$.  In comparison, the present-day
co-moving number density of galaxies brighter than L$^\ast$ in the B
band is $7\times10^{-4}~h^3_{50}$~Mpc$^{-3}$ (Ellis et al. 1996) and
the bright Lyman break galaxies at $z \sim 3$ reported by Steidel et
al. (1996) have a space density of $3.6\times
10^{-4}~h^3_{50}$~Mpc$^{-3}$.  The duty cycle of strong H$\alpha$
emission in galaxies with bursting or variable star formation histories
can be quite low, given the sensitivity of strong line emission to only
the most recent star formation. Thus it is not unreasonable that our
space density is below that of present day L$^*$ galaxies and
comparable to that of the UV continuum selected galaxies at high
redshift. The low resolution of the NICMOS grisms imprints a strong
selection effect against emission-lines with low equivalent widths.
Given the incompleteness at low equivalent widths and the rather bright
H-band magnitudes of the detected emission-line objects it seems
likely that we are sampling star formation in a population that is
closely related to the majority population of massive galaxies.

The average star formation rate in our sample is 30~M$_{\odot}$
yr$^{-1}$ (40~M$_{\odot}$ yr$^{-1}$ for ($q_0=0.1$).
After correcting for [NII] contamination, this rate is
reduced to 21~M$_{\odot}$ yr$^{-1}$.  We have not applied any
extinction correction to the calculations of H$\alpha$ luminosities and
star formation rates.  While our sample is not magnitude limited, the
distribution of apparent magnitudes is sufficiently narrow to allow a
reasonable comparison with magnitude limited samples. As noted above,
our median F160W magnitude of 20.5 is quite close to M$^{*}_R$ as
derived locally. It is interesting to note that the average star
formation rate in the Lyman break objects, before corrections for
extinction are $\sim 8$ M$_{\odot}$ yr$^{-1}$, and that the extinction
corrected rates are a factor of $3 - 7$ higher.  Recent near-infrared
spectroscopy of 8 CFRS galaxies at $z \sim 1$ by Glazebrook et al.
(1998) reached an H$\alpha$ luminosity of $2\times 10^{42}$erg/s/cm$^2$
with a star formation rate of $22 M_\odot$~yr$^{-1}$, comparable to
what we find, although our survey covers a wider range of H$\alpha$
luminosities and star formation rates.

One source of uncertainty in using our data to derive global star
formations rates is the unknown degree of contamination from AGNs. The
fraction of AGNs in our sample could, in principle, be calculated using
the AGN luminosity function in the redshift range from 0.7 to 1.9.
However, at z$=1.5$ an H magnitude of 20.5 corresponds roughly to an
absolute B magnitude of -22.5 adopting a B$-$H color similar that for
QSOs observed by Neugebauer et al. (1986). At this faint magnitude,
the quasar luminosity function is completely unknown.  The most recent
quasar luminosity function at z $\sim 0.5 - 2.0$ reaches only to M$_B
\sim -25$ (Hewett, Foltz \&\ Chaffee 1993).  Furthermore, our knowledge
of the Seyfert luminosity function is entirely based on local samples
and the degree of luminosity evolution in the Seyfert galaxy
populations as a function of redshift is unknown.  A crude estimate of
the fraction of AGNs in faint redshift surveys can be derived from
spectrophotometry of the CFRS sample. Using various emission line
ratios, Hammer et al. (1997) and Tresse et al. (1996) estimated that
about 8$-$17\%\ of their galaxies at $z < 0.7$ are active galaxies,
either Seyfert 2 or LINERs. A contamination rate at this level is
consistent with our classification of two of our objects as candidate
AGNs on the basis of their large equivalent widths and compact
morphologies. If we exclude the two AGN candidates, the average
SFR is 28.5~M$_\odot$/yr and 20~M$_\odot$/yr if we correct the
[NII]~6583,6548\AA\ contamination. The impact of line misidentifications is
comparable to that of the AGNs, as the most likely contaminant are [OIII]5007,4959
emission from compact non-stellar sources.

\bigskip
\section{Summary}
\bigskip

 We have presented the basic data derived from a spectroscopic survey
 of field galaxies in the near infrared. The G141 grism on NICMOS
provides a unique opportunity to survey large volumes to faint flux
levels in H$\alpha$ and other emission-lines. Our survey of 64 square
arcminutes at high and intermediate galactic latitudes resulted in the
detection of 33 emission-line objects and we believe that most or all
of these are H$\alpha$ at redshifts between 0.75 and 1.9.  The line
luminosities, equivalent widths and continuum magnitudes suggest that
we are seeing active star formation in galaxies from the upper end of
the luminosity function.  The apparent space density of H$\alpha$
emitters in our survey is similar to that of the bright Lyman break
objects identified at $z > 3$ and is within a factor of  $\sim 2$ of
that of present day L$^*$ galaxies. The grisms on NICMOS provide a
first look at the enormous potential of low resolution near-IR
spectroscopy from space.

\section{ Acknowledgments}
We thank the staff of the Space Telescope Science Institute for their
efforts in making this parallel program possible. In particular we
thank Peg Stanley, Doug van Orsow, and the staff of the PRESTO
division. We also thank John Mackenty and members of the STScI  NICMOS
group for crafting the exposure sequences.  We also acknowledge the
role of Duccio Macchetto and the Parallel Working Group, lead by Jay
Frogel, in making the public parallel program a success.  This research
was supported, in part, by grants from the Space Telescope Science
Institute, GO-7498.01-96A and P423101.  HIT and JPG acknowledge funding
by the Space Telescope Imaging Spectrograph Instrument Definition Team
through the National Optical Astronomy Observatories and by the NASA
Goddard Space Flight Center. \bigskip

\eject

\eject

\begin{figure}
\plotfiddle{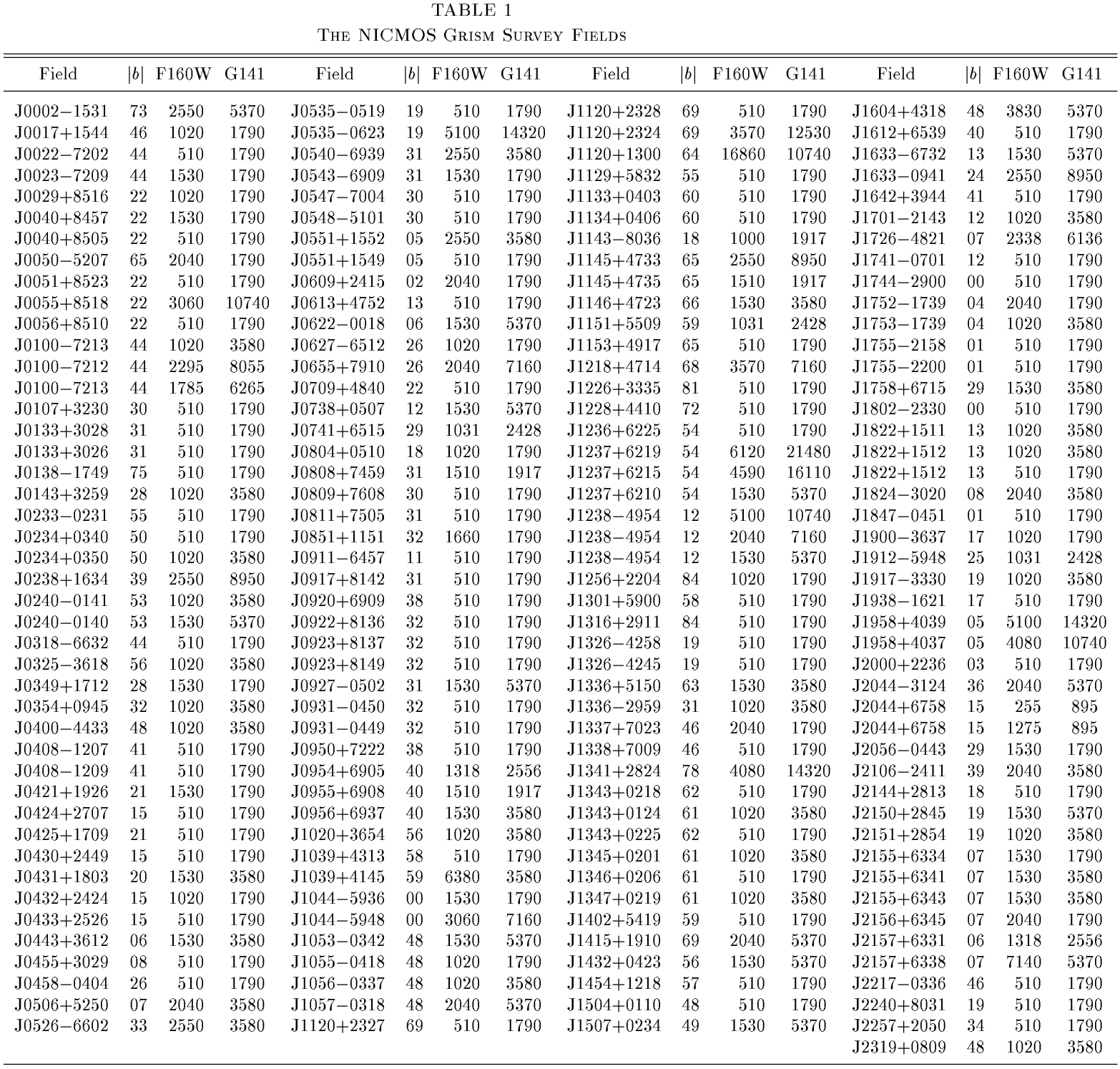}{12truecm}{0.0}{90}{100}{-310}{-250}
\end{figure}

\begin{figure}
\plotone{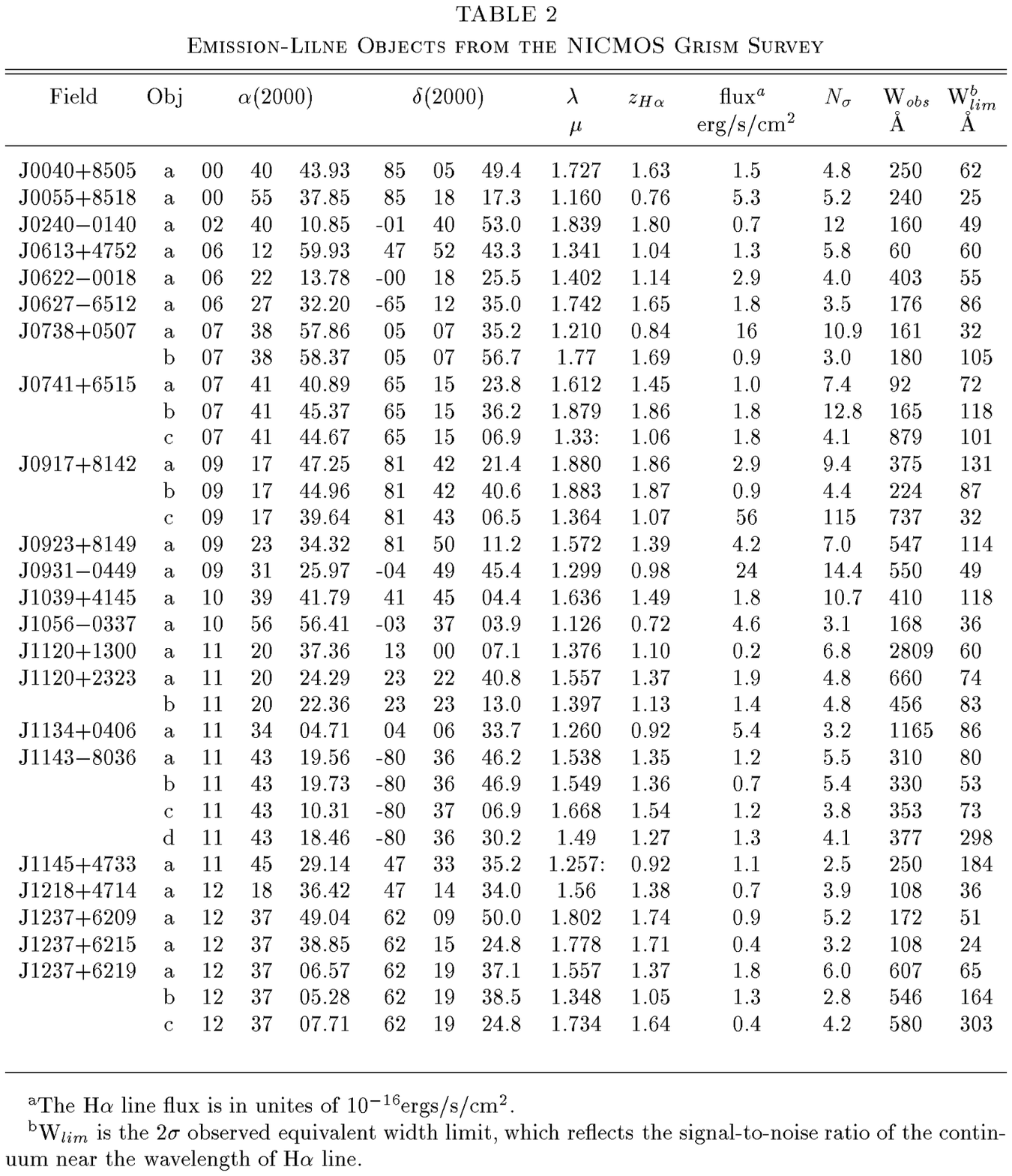}
\end{figure}

\begin{figure}
\plotone{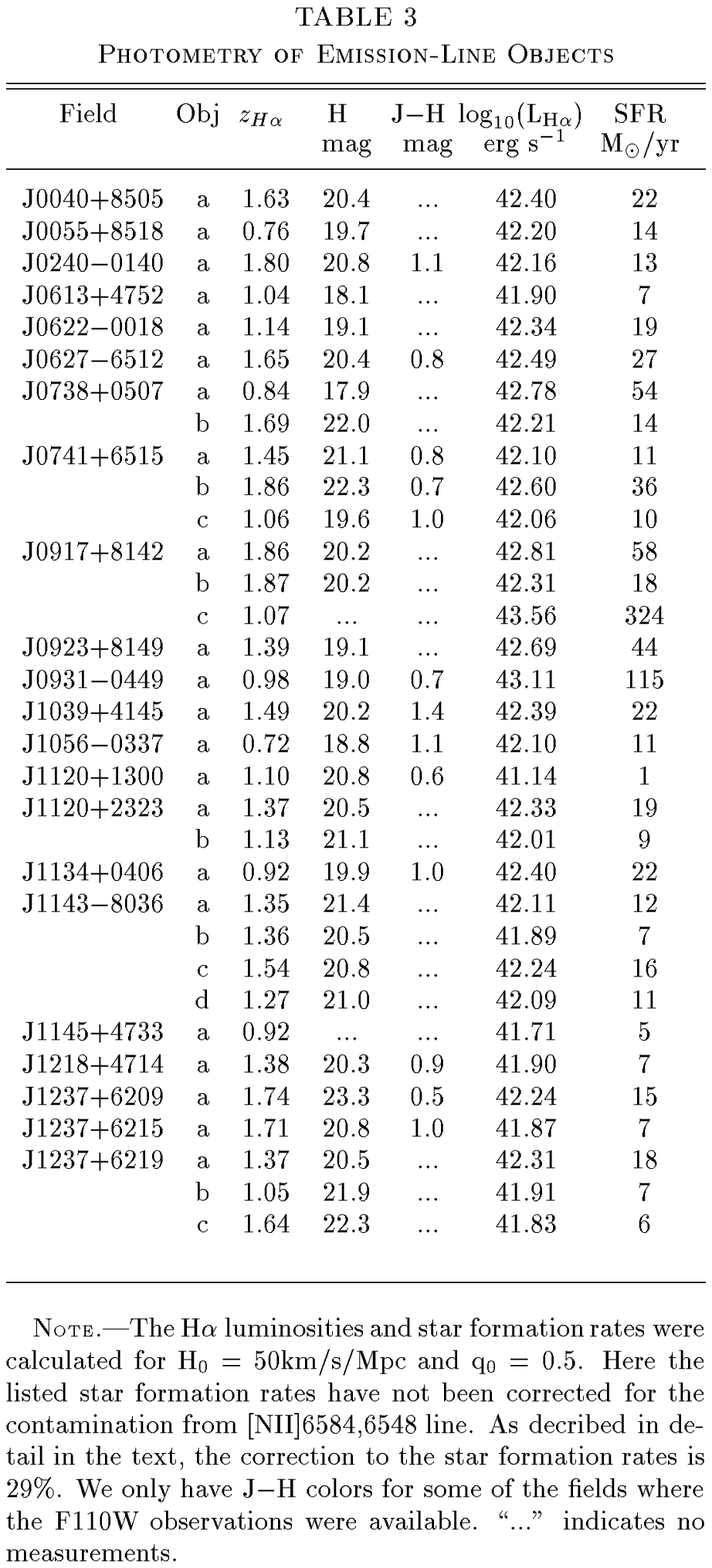}
\end{figure}

\begin{figure}
\plotone{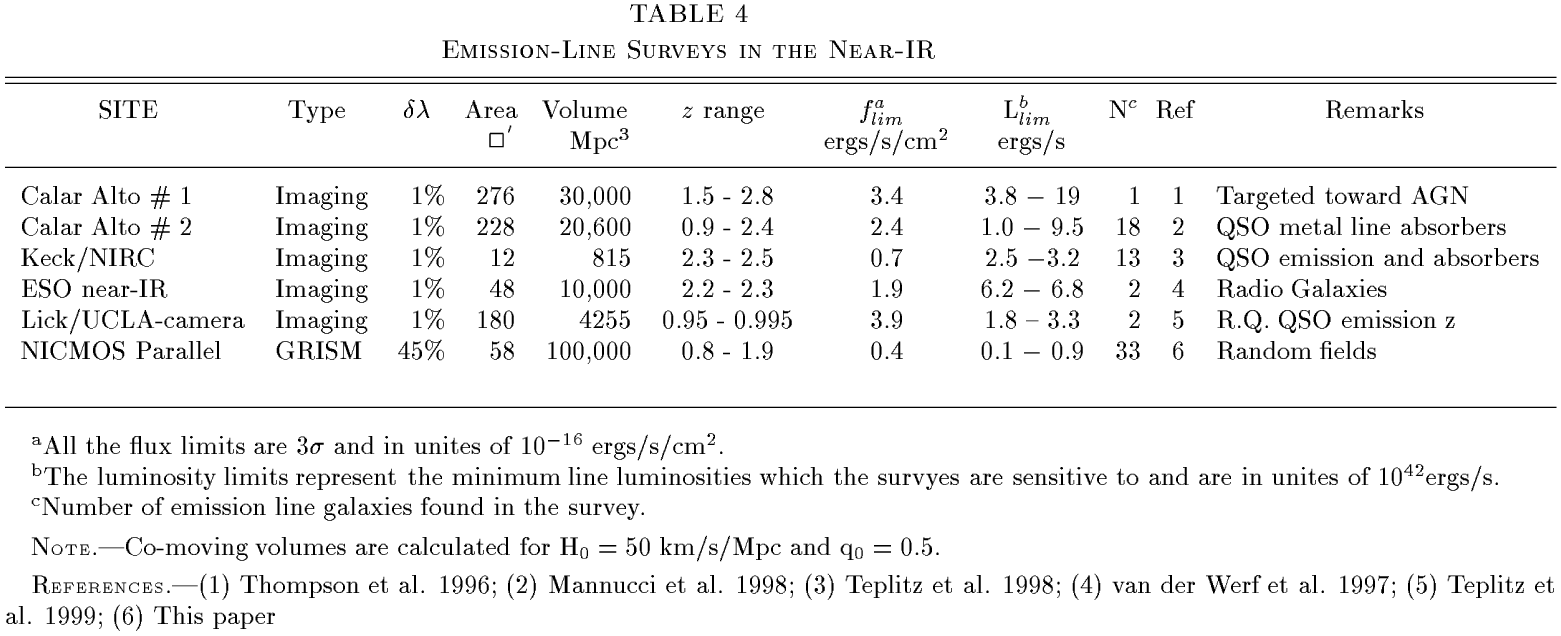}
\end{figure}


\begin{thebibliography}
\bigskip


\bibitem{}{Bechtold, J., Elston, R., Yee, H. K. C., Ellingson, E., \& Cutri, R. 1998,
in The Young Universe, ed. S. D'Odorico, A. Fontana, \& E.
Giallongo, ASP Conference Series, 146, 241}

\bibitem{}{Beckwith, S. V. W., Thompson, D., Mannucci, F., Djorgovski, S. G.
1998, ApJ, 504, 107}



\bibitem{}{Dey, A., Spinrad, H., Stern, D., Graham, J.R., Chaffee, F. 1998, ApJ, 498, 93L}

\bibitem{}{Dickinson, M.  1998, in The Hubble Deep Field", eds. M. Livio, M. Fall, and
P. Madau, STScI, in press}

\bibitem{}{Freudling, W. \&\ Pirzkal, N. 1998, in NICMOS and the VLT, eds. Wolfram Freudling and
Richard Hook, p47}

\bibitem{}{Ellis, R., Colless, M. Broadhurst, T. Heyl, J. \&\ Glazebrook, K., 1996, MNRAS, 280, 235}

\bibitem{}{Gallego, J., Zamorano, J., Rego, M., Vitores, A.G., 1997, ApJ, 475, 502}

\bibitem{}{Giavalisco, M., et al. 1998, in preparation.}

\bibitem{}{Glazebrook, K., Blake, C., Economou, F., Lilly, S.
\&\ Colless, M. 1998, astro-ph/9808276}

\bibitem{}{Hammer, F., Flores, H., Lilly, S.J., Crampton, D., Le Fevre, O., Rola, C., 
Mallen-Ornelas, G., Shade, D. \&\ Tresse, L. 1997, ApJ, 481, 49}

\bibitem{}{Heckman, T., Robert, C., Leitherer, C., Garnett, D.R., van der Rydt, F. 1998, ApJ,
503, 646}

\bibitem{}{Hewett, P., Foltz, C.B. \&\ Chaffee, F.H. 1993, ApJ, 406, L43}

\bibitem{}{Hu, E.M., Cowie, L.L. \&\ McMahon, R.G. 1998, ApJ, 502, L99}


\bibitem{}{Kennicutt, R.C., Bothun, G.D. \&\ Schommer, R.A. 1984, AJ, 89, 1279}

\bibitem{}{Kennicutt, R.C. 1992, ApJ, 388, 310}

\bibitem{}{Le Fevre, O., Hudson, D., Lilly, S.J., Crampton, D., Hammer, E. \&\ Tresse, L. 1996. ApJ, 461, 534}

\bibitem{}{Lilly, S.J., Le Fevre, O., Hammer, F. \&\ Crampton, D. 1996, ApJ, 
460, L1}

\bibitem{}{Lin, H., Kirshner, R.P., Shectman, S.A., Landy, S.D., Oemler, A., Tucker, D.L., Schechter, P.L. 1996, 464, 60}


\bibitem{}{Malkan, M. A., Teplitz, H., McLean, I. S. 1996, ApJ, 468, 9}

\bibitem{}{Mannucci, F., Thompson, D.J., Beckwith, S.V.W., Williger, G.M., 1998, ApJ, 501, 11}

\bibitem{}{Meurer, G.R., Heckman, T.M., Lehnert, M.D., Leitherer, C. \&\ Lowenthal, J. 1997,
 AJ, 114, 54}

\bibitem{}{Pettini, M., Kellogg, M., Steidel, C.C.,
Dickinson, M., Adelberger, K.L., Giavalisco, M. 1998, ApJ, in press}

\bibitem{}{Smail, I., Ivison, R.J., Blain, A.W. \&\ Kneib, J.P. 1998, ApJL, 507, L21}

\bibitem{}{Spinrad, H., Stern, D., Bunker, A., Dey, A., Lanzetta,
K., Yahil, A., Pascarelle, S. \& Fern\'andez-Soto, A. 1999, AJ, in press}
 
\bibitem{}{Steidel, C., Giavalisco, M., Pettini, M., Dickinson, M., Adelberger, K. 1996, ApJ, 462, L17}

\bibitem{}{Teplitz, H. I., Malkan, M. A., McLean, I., S. 1998, ApJ, in press}

\bibitem{}{Teplitz, H. I., McLean, I., S., \& Malkan, M. A. 1999a, ApJ, submitted }

\bibitem{}{Teplitz, H. I. Malkan, M. A., \& McLean, I. S. 1999b, in ``After The Dark Ages'', in press}

\bibitem{}{Thompson, D.J., Mannucci, F. \&\ Beckwith, S.V.W. 1996, AJ, 112, 1794}

\bibitem{}{Thompson, R.I., Rieke, M., Schneider, G., Hines, D.C., Corbin, M.R. 1998, ApJ, 49}

\bibitem{}{Tresse, L., Rola, C., Hammer, F., Stasinska, G., Le Fevre, O., Lilly, S.J. \&\ Crampton, D.
1996, MNRAS, 281, 847}
 
\bibitem{}{Weymann, R., Stern, D., Bunker, A., Spinrad, H., Chaffee, F., Thompson, R.I., 
Storrie-Lombardi, L.J. 1998, ApJ, 505, L95}

\bibitem{}{van der Werf, P.P., Bremer, M. N., Moorwood, A. F. M., Rottgering,
H. J. A. 1997, IAU, 186, 197}

\bibitem{}{Yan, L., McCarthy, P. J., Freudling, W., Teplitz, H. I., Malumuth, E. M.,
Weymann, R. J., Malkan, M. A., 1999, submitted to ApJ, Letters. }

\bibitem{}{Yee, H.K.C., Ellingson, E. \&\ Carlberg, R.G. 1996, ApJS, 102, 269}

\end{thebibliography}
\end{document}